%% file: main.tex
\documentclass[sigconf]{acmart}
\AtBeginDocument{%
  }

\usepackage{booktabs}
\usepackage{multirow}
\usepackage{makecell}
\usepackage{longtable}

\copyrightyear{2026}
\acmYear{2026}
\setcopyright{cc}
\setcctype{by}
\acmConference[RecSys '26]{20th ACM Conference on Recommender Systems}{September 27-October 02, 2026}{Minneapolis, MN, USA}
\acmBooktitle{20th ACM Conference on Recommender Systems (RecSys '26), September 27-October 02, 2026, Minneapolis, MN, USA}
\acmDOI{10.1145/3773078.3831818}
\acmISBN{979-8-4007-2284-4/2026/09}

\begin{document}

\title{On the Convergent Validity of Offline Evaluation Designs for Recommender Systems}

\author{Sushobhan Parajuli}
\email{sp3886@drexel.edu}
\orcid{0009-0008-5679-9524}
\affiliation{%
  \institution{Dept. of Information Science \\ Drexel University}
  \city{Philadelphia}
  \state{PA}
  \country{USA}
}

\author{Samira Vaez Barenji}
\email{svaez@drexel.edu}
\orcid{0000-0002-2123-4338}
\affiliation{%
 \institution{Dept. of Information Science \\ Drexel University}
  \city{Philadelphia}
  \state{PA}
  \country{USA}
}

\author{Michael D. Ekstrand}
% \correspondingauthor
\email{mdekstrand@drexel.edu}
\orcid{0000-0003-2467-0108}
\affiliation{%
 \institution{Dept. of Information Science \\ Drexel University}
  \city{Philadelphia}
  \state{PA}
  \country{USA}
}

\begin{abstract}
Offline evaluation on historical interaction logs is the most common evaluation methodology for recommender systems. 
However, such evaluations depend on sparse, incomplete, or biased data, which raises concerns about whether commonly used evaluation setups reliably reflect true user preferences.
In this work, we study how offline evaluation design choices affect the validity of recommender system comparisons.
We evaluate a set of recommendation models across several evaluation setups that vary key factors such as data filtering thresholds and candidate set construction.
To assess the validity of these configurations, we measure the correlation between model rankings obtained from conventional train–test splits on sparse interaction data and rankings from evaluations based on dense ground-truth user feedback.
We use this agreement as an indication of their validity with respect to true user preferences.
Our results show that the validity of sparse evaluation depends on the dataset and the specific dense evaluation targets, and that there is no uniformly best offline evaluation design.
\end{abstract}




\maketitle

\section{Introduction}
Offline evaluation based on historical interaction logs is the dominant approach for developing and comparing recommender systems (RS) in academic research, and serves as the first evaluation step in industrial settings where promising algorithms undergo further online trials.
By splitting observed user–item interactions into training and test sets, researchers can efficiently evaluate models without requiring costly user studies or online experiments.

However, despite their widespread use, offline evaluation methods have significant limitations in how well they reflect true user preferences or estimate the likely online effectiveness of a proposed system.
These limitations include \emph{reliability} concerns, where results are heavily dependent on specific experimental setups or random seeds; \emph{internal validity} concerns, caused by challenges such as extreme data sparsity and the biased nature of interaction data, where metrics may fail to capture what they are intended to measure; and \emph{external validity} concerns, where conclusions do not necessarily transfer to other settings such as online deployment.
Therefore, conclusions drawn from offline experiments may not faithfully reflect the underlying construct of interest, such as user satisfaction or system effectiveness.

In this paper, we examine challenges to the internal validity of offline top-N recommender system evaluation practices that arise from the extreme sparsity and biased observations in recommender system datasets.
A key challenge is that we only observe feedback for a small subset of user–item pairs in these datasets, and that the missing interaction data is missing not at random (MNAR) \citep{marlinCollaborativePredictionRanking2009a, steckTrainingTestingRecommender2010, marlinCollaborativeFilteringMissing2012}.
These interactions cannot be reliably interpreted as either negative or unknown feedback, making it difficult to obtain valid measurements of model performance.

Offline evaluation results are further influenced by several experimental design choices, including data filtering, candidate set construction, and train–test splitting strategies.
Prior work shows that these choices can substantially affect the ranking of recommendation models \citep{gusakTimeSplitExploring2025, mengExploringDataSplitting2020, kricheneSampledMetricsItem2020, pereiraReliabilitySamplingStrategies2025}, but it remains unclear which configurations most reliably reflect true user preferences.
We therefore evaluate how well different evaluation setups on sparse data predict model rankings obtained from dense ground-truth feedback.

Using recently-released datasets that include relatively small but dense sets of user feedback, we assess the \emph{convergent validity} between sparse top-N evaluations with sparse data and comparable evaluations using dense user feedback.
Convergent validity, a concept from measurement theory \cite{salaudeenMeasurementMeaningValidityCentered2025, wallachPositionEvaluatingGenerative2025, wangEvaluatingGeneralPurposeAI2026}, is a specific aspect of internal validity that is achieved when two different measurement procedures intended to capture the same construct produce consistent results.
In our setting, this corresponds to comparing model rankings obtained from a commonly used, inexpensive measurement procedure (sparse offline evaluation) with rankings derived from a higher-quality but more costly alternative (dense ground-truth evaluation).
Our study is guided by two key questions:
\begin{description}
    \item[RQ1] To what extent do model rankings obtained from sparse offline evaluation correlate with rankings derived from dense ground-truth evaluation?
    \item[RQ2] Do modifications to offline evaluation, such as sampling and data filtering, reduce the gap between these evaluations?
\end{description}

To address these questions, we evaluate a diverse set of recommendation models across a range of evaluation configurations, varying data filtering, candidate set construction, and list length.
For each configuration, we evaluate the models on the sparse data and measure the correlation between those results and evaluation on the dense data.
Our results suggest that the agreement between sparse and dense evaluation depends strongly on the dataset and evaluation setup, and that commonly observed differences between recommendation models may be driven as much by evaluation protocols as by model effectiveness.

\section{Background and Related Work}
\paragraph{Limitations of offline evaluation.}
A large body of prior work has examined the limitations of offline evaluation in RS, particularly its ability to reflect real-world performance and user preferences \citep{hidasiWidespreadFlawsOffline2023, jadidinejadSimpsonsParadoxOffline2022, jeunenRevisitingOfflineEvaluation2019}.
Rankings of algorithms derived from offline metrics often fail to generalize to user-facing settings, raising concerns about the external validity of standard evaluation protocols \citep{rossettiContrastingOfflineOnline2016, beelComparisonOfflineEvaluations2015}. 
Part of this gap is fundamental: even an ideal offline evaluation with perfect and complete or unbiased data may not fully capture interface effects, feedback loops, or long-term user satisfaction.
That is, even a perfect offline evaluation may lack external validity.
However, there is also a gap between a hypothetical ideal offline evaluation and what can be measured within the practical reality of sparse, biased interaction data and simplified evaluation protocols: offline evaluation in practice lacks internal validity with respect to what it purports to measure.
We focus on the second component by examining how much evaluation design choices contribute to the gap between practical and ideal.

One major contributor to the internal validity gap is the nature of the data used for RS evaluation. Interaction data is typically sparse and missing not at random (MNAR), meaning that observed interactions are influenced by exposure and user behavior rather than random sampling \citep{marlinCollaborativePredictionRanking2009a, steckTrainingTestingRecommender2010}.
Therefore, missing interactions can carry useful information about user's taste.
\citet{steckTrainingTestingRecommender2010} shows that treating missing interactions as unknown or randomly missing can introduce systematic bias in both training and evaluation. 
Later work proposed sampling techniques to attempt to approximate missing-at-random conditions or correct for exposure bias \citep{carraroSamplingApproachDebiasing2022}.

\paragraph{Evaluation as measurement.}
Measurement theory provides a number of validity and reliability properties for assessing the quality of empirical measurements \citep{kimberlinValidityReliabilityMeasurement2008, cookCurrentConceptsValidity2006}.
These properties provide a framework to more precisely characterize \emph{how} internal or external validity may be violated.
In this paper, we are particularly concerned with \textbf{convergent validity}: the extent to which multiple measurements intended to capture the same construct produce consistent results.
Specifically, we compare offline evaluation results with sparse data to those obtained from dense ground-truth feedback to assess the extent to which these two measurements of model effectiveness converge on consistent answers.
% We adopt convergent validity as a lens to study offline evaluation in RS by comparing model rankings derived from sparse sparse interaction data with those obtained from dense ground-truth feedback.
% Our analysis can also be framed as a \emph{predictive validity} problem, which considers whether a measurement procedure can accurately predict outcomes observed under a more reliable or realistic setting, in our case, evaluation results based on dense preference data that are closer to true user preferences.

\paragraph{Evaluation design choices.}
Experimental design choices in offline evaluation can significantly affect the outcome of model comparisons.
For example, different train–test splits, such as random, leave-one-out, or temporal splits, can lead to substantially different rankings of recommender systems \citep{gusakTimeSplitExploring2025, mengExploringDataSplitting2020}.
\citet{sunTakeFreshLook2023} and \citet{jiCriticalStudyData2023} show that ignoring temporal ordering can introduce leakage and unrealistic evaluation scenarios.
\citet{verachtertrobinAreWeForgetting2022} suggest using an ``optimal training window'' of only recent data can drastically improve performance and change model rankings compared to using entire static datasets, as models otherwise learn patterns no longer aligned with current user interests; however, some forms of data pruning like $k$-core filtering can remove large portions of users and make results appear better by excluding cases on which algorithms perform worse \citep{Beel2019e}.

Another offline evaluation decision of particular interest with respect to evaluation data selection bias is \emph{candidate set sampling}: restricting the recommender to using a random subset of the items as candidates.
While this has often been done for performance reasons \cite{korenFactorizationMeetsNeighborhood2008, cremonesiPerformanceRecommenderAlgorithms2010}, it has also been proposed as a possible strategy to mitigate data biases \cite{ekstrandSturgeonCoolKids2017, ihemelanduCandidateSetSampling2023, belloginStatisticalBiasesInformation2017}.
\citet{ihemelanduCandidateSetSampling2023} used simulation to find that weighting candidate sampling by item popularity may help reduce evaluation bias.
\citet{canamaresTargetItemSampling2020} found that minimum target sets can produce conclusions that differ from evaluation over the full candidate set, while neither minimal nor maximal target sets are uniformly optimal.

One of the persistent difficulties in evaluating evaluation designs is finding a reference point.
Observing that two methods produce different results does not, on its own, tell us which method is more valid or reliable.
For example, \citet{kricheneSampledMetricsItem2020} concluded that sampled candidate sets produce biased evaluations by comparing their results against full-set evaluations and finding significant differences in outcomes.
However, if the sampling compensates for a bias in the original data, it may be more correct even though it differs.
% A difference in outcome, however, is not sufficient to conclude \emph{which} outcome is correct: if sampling effectively corrects for a bias present in unsampled evaluation, .
% They compare sampled and full metrics on the same sparse data which implies that if full evaluation on sparse implicit data does not reflect true user preferences, then consistency between sampled and full evaluation on that data is not sufficient to establish the validity of the evaluation protocol.
In this paper, we address this challenge by using smaller, harder-to-collect dense ground-truth data as a reference point for comparing evaluation designs over traditional sparse data.

\begin{table}
\caption{Models employed.}
\label{tbl:models}
\begin{minipage}{\columnwidth}
\centering\small
\input{tables/models_table}
\end{minipage}
\end{table}

\begin{figure}
    \centering
    \includegraphics[width=\columnwidth]{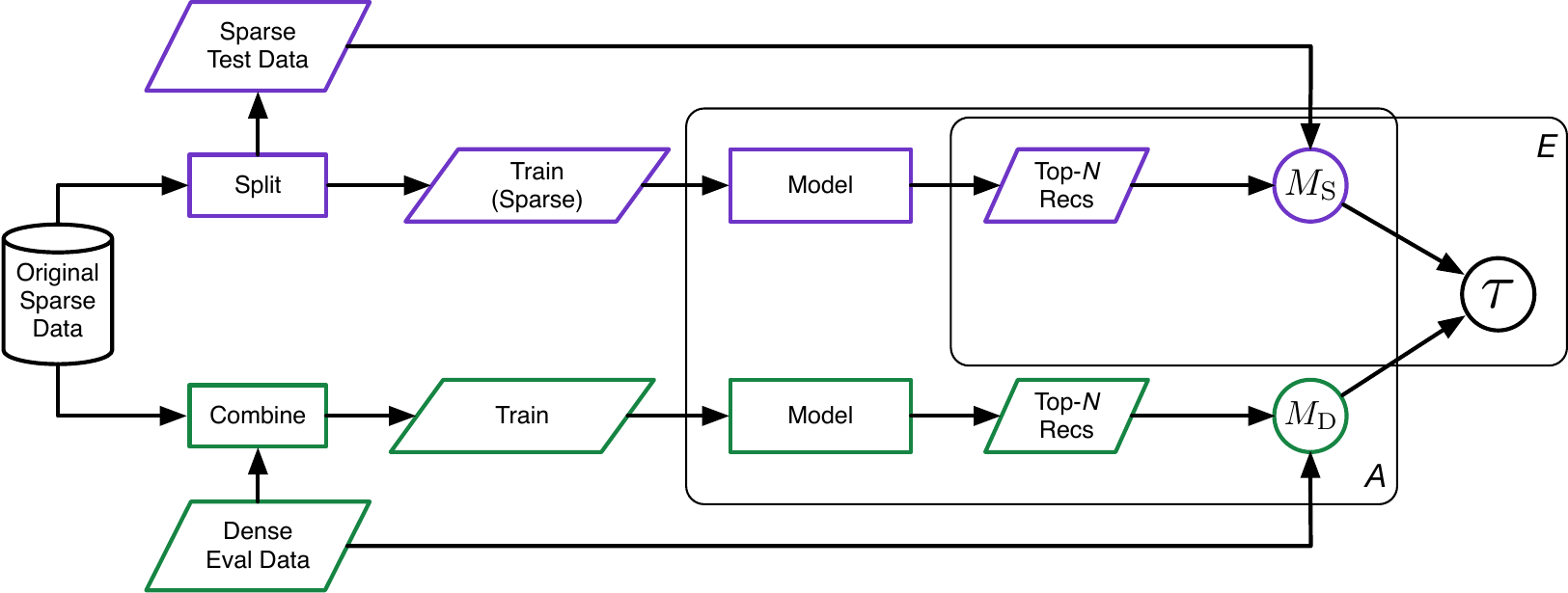}
    \caption{Diagram of experimental design for a single dataset and dense eval target.  $M_{\mathrm{S}}$ and $M_{\mathrm{D}}$ are evaluation
    metrics over sparse and dense data, respectively, and the plates indicate repetition for $A$ models and $E$ sparse evaluation designs.}
    \label{fig:experiment}
\end{figure}

\section{Experimental Design}

The core of our experimental design is to compare two evaluation setups applied to the same set of recommendation models, shown in Figure~\ref{fig:experiment}.
In the \textbf{sparse setup}, we split the sparse data into training and test sets.
We then train and evaluate the models using a typical offline evaluation setup.
In the \textbf{dense setup}, we train models on the entire sparse data and evaluate them against a dense ground-truth collected through an experimental intervention, in which users are exposed to many or all items to obtain much more complete preference data than is typically available.
Each dataset therefore consists of two components: sparse data and dense ground-truth constructed through this intervention.
The precise form of the dense ground-truth varies between datasets, and we describe it in more detail below.

We repeat the experiments on two different datasets (MovieLens and KuaiRec), a range of sparse evaluation design choices (candidate set construction, list length, and relevance threshold), and multiple dense evaluation targets.
For each, we measure how well the sparse evaluation's ranking of recommendation models correlates with the ranking under dense evaluation using Kendall's $\tau$. 
This allows us to assess the extent to which the sparse data available for offline recommender system evaluation is a valid proxy for dense ground truth (RQ1), and to identify which sparse evaluation design choices --- candidate set construction, list length, and relevance threshold --- best preserve that validity (RQ2).\footnote{The code is available at \url{https://doi.org/10.5281/zenodo.21549024}}

\subsection{Datasets}
\label{sec:data}
\paragraph{MovieLens-32M}
MovieLens-32M is a movie recommendation dataset containing 32 million explicit ratings from 200,948 users on a 0.5-5.0 scale \citep{harperMovieLensDatasetsHistory2015a}.
The extended version proposed by \citet{smuckerExtendingMovieLens32MProvide2025} augments this dataset with a dense set of relevance judgments collected from 51 active MovieLens users.
These users were recruited and asked to assess recommended movies by indicating their interest in watching them, providing explicit preference judgments beyond their existing rating histories.
To collect these judgments, recommendation lists from multiple algorithms were pooled together following standard information retrieval pooling methodology \cite{voorhees2005trec, luEffectPoolingEvaluation2016}.
Participants then evaluated the pooled items to produce a denser set of user–item relevance signals than what is available in sparse rating data.

The resulting dataset contains approximately 31.7 million ratings on the standard MovieLens 0.5-5.0 scale.
We use this dataset as our sparse data for evaluation and as the training data for dense evaluation.
For the dense ground-truth we use \textit{interest.qrels} from the extension, which captures each participant's reported interest level in watching each movie on a 0--4 scale.
We collapse the original 0--7 scale to 0--4 by merging the top ranking levels.
For binary evaluation we threshold at interest level $r\ge 2$ (Interested) and $r\ge 3$ (Very Interested). 
For graded evaluation we use the 0--4 interest level as-is.

\paragraph{KuaiRec}
KuaiRec is a short-video recommendation dataset collected from the Kuaishou platform \citep{gaoKuaiRecFullyobservedDataset2022}.
The dataset contains user interactions based on watch behavior, where the primary feedback signal is the \emph{watch ratio}, defined as the play duration divided by video duration.
A watch ratio of 1.0 indicates the user watched the video at least once in full, and greater than 2.0 indicates the video was replayed.
When using explicit models, we treat watch ratios as ratings.

KuaiRec provides a ``big'' matrix consisting of 7,176 users and 10,728 items at 16.3\% density.
We use this matrix as our sparse data for evaluation and as training data for dense evaluation.
For dense ground-truth, we use the ``small'' matrix provided in KuaiRec, which contains 1,411 users and 3,327 items at 99.6\% density.
This matrix was collected through a controlled exposure process in which users were shown a large portion of the item catalog \citep{gaoKuaiRecFullyobservedDataset2022}.
For binary evaluation we threshold at watch ratio $r\ge 0.85$ and $r\ge 2.0$.
For graded evaluation we subtract 0.2 from the watch ratio and clip at zero, so videos watched less than 20\% through receive zero gain.

% All users and items in the small matrix are present in the big matrix, but the interactions in the two matrices are disjoint.
% This design enables training on sparse interaction from the big matrix and evaluating on a densely observed set of user-item interactions.

\subsection{Evaluation Setup}
To generate the model rankings used in our comparison, we evaluated 45 recommendation models from LensKit 2026.1.0 \citep{ekstrand2020lenskit}, including both implicit and explicit feedback models.
Table~\ref{tbl:models} lists the models and their variants.
Explicit models were trained directly on the original feedback values.
For implicit models, we additionally vary the minimum feedback value required for an interaction to be included in training --- three cutoffs for MovieLens and four for KuaiRec.
This yields 97 model configurations for MovieLens and 123 for KuaiRec, where each configuration corresponds to a distinct model and training threshold combination.

\paragraph{Sparse}
In the sparse evaluation, we train models on the sparse data and evaluate them on a held-out test set from the same dataset.
For MovieLens we use a temporal split with a fixed cutoff date so all training interactions occur before test interactions.
The data in the KuaiRec matrices do not follow temporal order, so we use 5-fold cross-validation, holding out 5 interactions per user per fold.

We study three candidate set conditions: full candidates (all unseen items), U-1000 candidates (1,000 uniformly sample items), and P-1000 candidates (1,000 popularity-weighted sample items).
We evaluate using both binary and graded relevance.
For binary relevance, a test interaction is counted as relevant if the rating or watch ratio meets or exceeds a threshold.
In the results tables, threshold rows reflect different ways of defining relevance at evaluation time, not different training configurations.
We compute $\mathrm{NDCG}@k$ and $\mathrm{DCG}@k$ at ranking cutoffs $k \in \{10, 20, 100\}$.

\paragraph{Dense}
In this setup, we train models on the sparse data and evaluate them against the dense ground-truth data described in Section~\ref{sec:data}.
For MovieLens and KuaiRec, this corresponds to evaluation on the pooled user judgments (interest.qrels) and the small dense matrix, respectively.
We evaluate using both binary and graded relevance and compute $\mathrm{NDCG}@k$ and $\mathrm{DCG}@k$ at ranking cutoffs $k \in \{10, 100\}$.
Results at $k=20$ were comparable to those at $k=10$ and added little additional information, so we selected the two more widely separated cutoff values to represent distinct evaluation settings while keeping the result tables manageable.

\begin{table*}[bt]
\centering
\caption{Kendall's $\tau$ between sparse and dense evaluation model rankings by NDCG (MovieLens). Cells show $\tau$ with bootstrapped 95\% confidence intervals. \textbf{Bold} indicates highest $\tau$ per dense target column.}
\label{tab:hero_ml}
\resizebox{\textwidth}{!}{
\input{tables/hero_table_ml_ndcg}
}
\end{table*}

\begin{table*}[bt]
\centering
\small
\caption{Top-5 models by dense NDCG@10 (k=10, Very Interested) and their best rank under the best-correlated sparse design and their best rank under any sparse design (MovieLens). The best dense model's sparse rank is in \textbf{bold}.}
\label{tab:best_dense_rank_ml}
\input{tables/best_dense_rank_ml}
\end{table*}

\begin{table*}[bt]
\centering
\caption{Kendall's $\tau$ between sparse and dense evaluation model rankings by NDCG (KuaiRec). Cells show $\tau$ with bootstrapped 95\% confidence intervals. \textbf{Bold} indicates highest $\tau$ per dense target column.}
\label{tab:hero_kr}
\resizebox{\textwidth}{!}{
\input{tables/hero_table_kr_ndcg}
}
\end{table*}

\begin{table*}[bt]
\centering
\small
\caption{Top-5 models by dense NDCG@10 (k=10, Watched 2$\times$) and their rank under the best-correlated sparse design and their best rank under any sparse design (KuaiRec). The best dense model's sparse rank is in \textbf{bold}.}
\label{tab:best_dense_rank_kr}
\input{tables/best_dense_rank_kr}
\end{table*}

% \subsection{Meta-Evaluation}
% For each sparse evaluation, we compute rank correlation (using Kendall's $\tau$) with each of the dense evaluations.
% This measures how well each sparse evaluation predicts the relative performance of recommendation models when evaluated with the dense data.

\section{Results}
In this section, we examine the correlation between sparse and dense model rankings, and assess whether changes in evaluation design affect the gap between them.
Our primary results are from Kendall's $\tau$ correlations between sparse and dense evaluation rankings of recommendation models.
Each correlation is computed across the complete set of model configurations, including both explicit and implicit models.
Tables~\ref{tab:hero_ml} and ~\ref{tab:hero_kr} report Kendall's $\tau$ values.

\subsection{MovieLens}

Table~\ref{tab:hero_ml} presents Kendall's $\tau$ values between sparse and dense model rankings for MovieLens.
The dense target columns reflect two levels of user interest: Interested ($r\geq2.0$) and Very Interested ($r\geq3.0$), each at $k=10$ and $k=100$.
We focus on Very Interested at $k=10$ as the most practically meaningful target --- identifying models that surface movies users are genuinely interested in watching within a short list --- and use the other targets to contextualize the findings.

\begin{figure}
    \centering
    \includegraphics[width=\columnwidth]{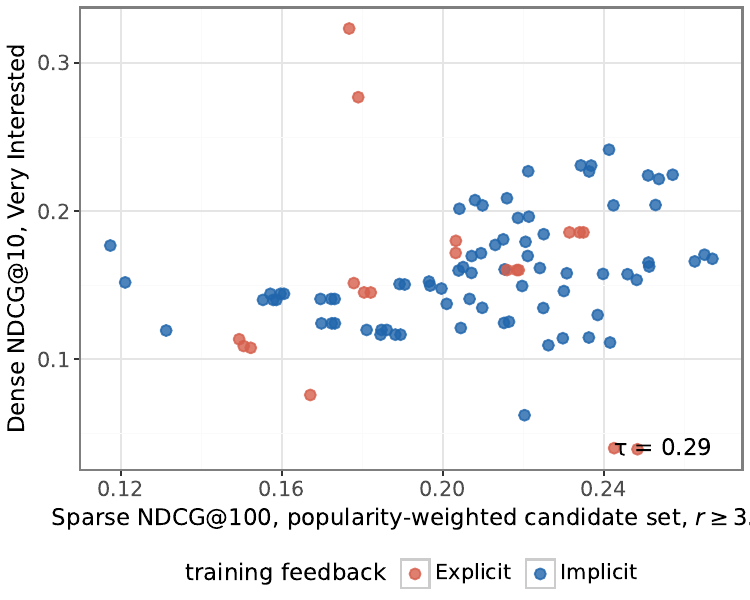}
    \caption{Scatter plot of NDCG scores under sparse evaluation (popularity-weighted candidate set, $k=100$, $r\geq$3.0) versus dense evaluation ($k=10$, Very Interested) on MovieLens.}
    \label{fig:ml_plot_pop}
\end{figure}

\begin{figure}
    \centering
    \includegraphics[width=\columnwidth]{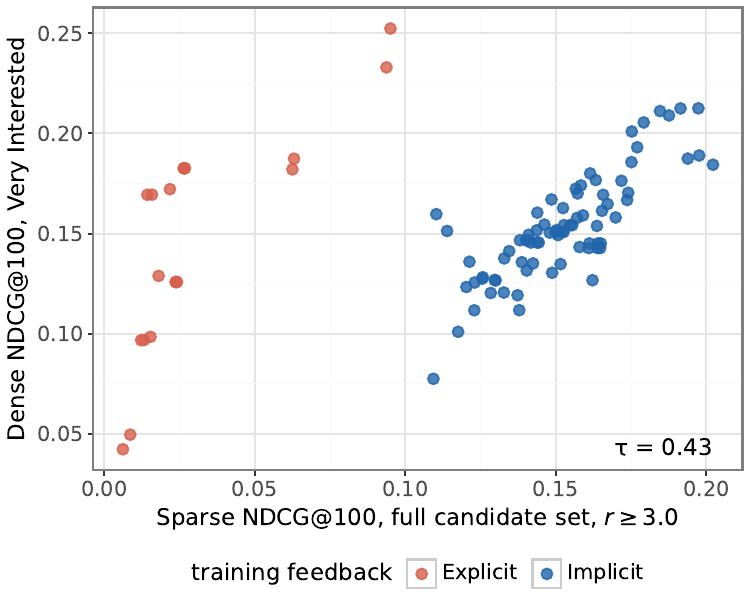}
    \caption{Scatter plot of NDCG scores under sparse evaluation (full candidate set, $k=100$, $r\geq$3.0) versus dense evaluation ($k=100$, Very Interested) on MovieLens.}
    \label{fig:ml_plot_full}
\end{figure}

\paragraph{RQ1: Alignment between sparse and dense rankings.}
Sparse evaluation has positive correlation with dense evaluation under all designs in our selected dense target.
For the dense target Very Interested and $k=10$, the highest correlation is obtained with the sparse P-1000 candidate set at $k=100$ and threshold $r \geq 3.0$, with $\tau=0.29$. Figure~\ref{fig:ml_plot_pop} shows the relationship between model scores under these two evaluation configurations.
The P-1000 candidate set generally produces higher $\tau$ values than the other candidate set designs, although the differences are small.
For example, full candidates achieve $\tau=0.26$ on the same sparse $k$ and threshold.
The correlation improves when the dense target is extended to $k=100$, however the sparse design it improves on is a different one --- full candidate set at sparse $k=100$ and $r\geq3.0$. 
This sparse evaluation design achieves $\tau=0.43$, the highest value in the table.
This improvement from $k=10$ to $k=100$ dense target may reflect that sparse evaluation is better suited to predicting which models place relevant items somewhere in a long list than which models surface them at the very top.

Table~\ref{tab:best_dense_rank_ml} shows the best models in dense evaluation and their rankings in sparse evaluation.
Some of the top models that use implicit feedback in training --- fsSLIM variants --- rank consistently well under both sparse and dense evaluation.
However, the best dense models overall that use explicit feedback --- BiasedSVD-d64 and BiasedSVD-d32 --- rank only $23^{rd}$ and $22^{nd}$ under their best sparse design, and rank $40^{th}$ and $38^{th}$ under the best-correlated sparse design (Full, $k=100, r\geq3.0$).
When we compare the NDCG columns for these BiasedSVD variants, the metric values under dense evaluation are a lot higher than those under sparse evaluation.
So, the BiasedSVD models perform significantly better on dense evaluation but are misranked on sparse evaluation.
Similar ranking differences are visible for other models (see Figures~\ref{fig:ml_plot_pop} and ~\ref{fig:ml_plot_full}).
Models trained on implicit feedback tend to show more consistent performance across sparse and dense evaluations, whereas models trained on explicit feedback show greater variation.

\paragraph{RQ2: Impact of evaluation design choices.}
Among the design choices we consider, candidate set has the largest visible effect on rank correlation, followed by relevance threshold and sparse $k$.

Against the dense target Very Interested and $k=10$, the three candidate sets produce similar $\tau$ values --- full candidate set at sparse $k=10, r\geq3.0$, achieves $\tau=0.21$, U-1000 candidate set achieves $\tau=0.24$, and P-1000 candidate set achieves $\tau=0.28$.
P-1000 candidate set is consistently better than the other designs at this dense target.

Against the Very Interested and $k=100$ dense target, the candidate set effect becomes large and clear.
Full candidate set at sparse $k=100$, $r\geq3.0$ achieves $\tau=0.43$.
U-1000 candidate set at the same sparse design drops to $\tau=0.36$, and P-1000 candidate set drops further to $\tau=0.18$.
This reversal of P-1000 candidate set's relative performance between $k=10$ and $k=100$ dense targets suggests that popularity-weighted sampling may slightly favor models that perform well at short-list dense targets, but it does not improve alignment when the dense target is a longer list.
Across all dense targets at $k=100$ --- Interested, Very Interested, and Graded --- full candidate set designs produce consistently higher $\tau$ values.
This suggests that candidate set choice has the most visible impact on model ranking correlation against dense evaluation. 

Relevance threshold is a secondary factor.
Graded sparse evaluation underperforms binary threshold $r\geq3.0$, but slightly outperforms $r\geq4.0$ against most dense targets.
For example, full candidate set at sparse $k=100$ with graded relevance achieve $\tau=0.40$ against Very Interested and $k=100$, compared to $\tau=0.43$ with $r\geq3.0$ and $\tau=0.38$ with $r\geq4.0$.
This difference is small but consistent across full and U-1000 candidate sets.
However, this pattern is mostly absent for P-1000 candidate set.
Between binary thresholds, $r\geq3.0$ outperforms $r\geq4.0$ across full and U-1000 candidate sets. 
But, this pattern is reversed for P-1000 candidate set.

Sparse $k$ has a small effect on rank correlation.
$\tau$ values at $k=10, k=20$, and $k=100$ are nearly identical within each candidate set and threshold combination.
This suggests that the number of recommendations generated does not meaningfully affect how well sparse evaluation predicts dense rankings.

\subsection{KuaiRec}
While MovieLens showed weak but positive correlation between sparse and dense evaluations, our KuaiRec experiments find the evaluation designs to be negatively correlated (see Table~\ref{tab:hero_kr}).
For this dataset, sparse evaluation inverts the dense ground-truth ranking regardless of any design choice. 
The two evaluation designs are incompatible for KuaiRec, and no evaluation design choice resolves this issue.

\begin{figure}
    \centering
    \includegraphics[width=\columnwidth]{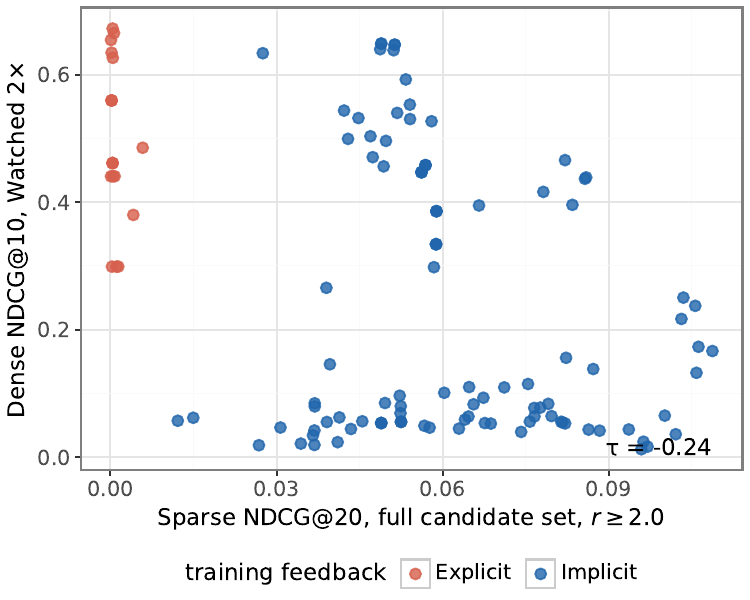}
    \caption{Scatter plot of NDCG scores under sparse evaluation (full candidate set, $k=20$, $r\geq$2.0) versus dense evaluation ($k=10$, Watched 2x) on KuaiRec.}
    \label{fig:kr_plot}
\end{figure}

\paragraph{RQ1: Alignment between sparse and dense rankings.}
Against the dense target Watched 2x and $k=10$, the least negative value is achieved by sparse P-1000 candidate set at $k=20, r\geq2.0$: $\tau=-0.19$.
Every other design produces more negative $\tau$.
Against the Watched $\geq85\%$ and $k=10$, the least negative $\tau$ is at sparse full candidate set, $k=20, r\geq2.0$. 
Against the dense target Graded, $\tau$ values are most extreme --- ranging to $-0.57$ --- which indicates that graded relevance signals amplify the inversion most severely.
Even under the most favorable sparse design, the top-5 dense models rank between $17^{th}$ and $28^{th}$ (see Table~\ref{tab:best_dense_rank_kr}).
Under the best-correlated sparse design (Full, $k=20$, $r\geq2.0$), ranks worsen further, ranging from $17^{th}$ to $44^{th}$.
Overall, sparse evaluation fails to reflect true model performance, as further illustrated in Figure~\ref{fig:kr_plot}, where models trained with explicit feedback perform poorly.

These findings from KuaiRec data show that, for this dataset, sparse evaluation can produce model rankings that are effectively inverted relative to dense ground-truth rankings.

\paragraph{RQ2: Impact of evaluation design choices.}
Among the design choices we consider, relevance threshold has the largest effect.
For the Watched 2x and $k=10$, sparse full candidates at $k=20$ achieve $\tau=-0.24$ at $r\geq2.0$.
However, for other sparse threshold designs under the same candidate set and $k$ the $\tau$ decreases to $-0.47$.
Sparse $k=20$ produces the least negative $\tau$ values across all candidate sets and dense targets.

Candidate set construction has a small effect on rank correlation of the models. Sparse P-1000 candidate set at $k=20$ and $r\geq2.0$ achieve $\tau=-0.19$ against the Watched 2x and $k=10$, compared to $\tau=-0.24$ for sparse full candidate set and $\tau=-0.21$ for sparse U-1000 candidate set.
However, given the confidence interval overlap we do not draw strong conclusions that changes in candidate set construction significantly impact model rank correlation.

\section{Discussion and Limitations}
Our results show that the correlation between sparse and dense evaluation varies by dataset and evaluation target. 
For MovieLens, there is positive correlation under most binary evaluation designs, while graded designs show lower correlations, with more negative values compared to binary evaluation designs. 
For KuaiRec, sparse evaluation inverts dense model rankings across all design choices, producing mostly negative correlations.

The contrast between the mostly positive MovieLens correlations and the consistently negative KuaiRec correlations could be partially attributed to their underlying characteristics, although we did not isolate the individual effects of such factors in our analysis.
First, watch ratios in KuaiRec are implicit signals, while ratings in MovieLens are explicit signals. 
As a result, explicit models may have performed worse in the sparse evaluation design, where watch ratios are treated as ratings.
Second, the KuaiRec dataset does not preserve temporal order, and the experiments rely on cross-validation as a non-temporal split, which may not reflect realistic recommendation scenarios.  
KuaiRec also contains fewer users than MovieLens, which may further affect the stability of evaluation results. 
For MovieLens, the positive correlations may be the effect of pooling bias and k-core filtering in the extended version that we used, which emphasize popular items and increase overlap between sparse and dense evaluation; we leave further analysis of this effect for future work.

We also observe that design modifications to sparse evaluation can affect alignment with dense evaluation.
For MovieLens, full candidate sets and binary thresholds tend to produce higher correlations than sampled candidate sets and graded evaluation.
These findings are consistent with prior work \citep{kricheneSampledMetricsItem2020}, in that sparse evaluation with full candidate sets is better correlated with dense evaluation than evaluation with uniformly sampled candidate sets.
Popularity-weighted sampling can improve correlation for some dense evaluation targets, but that improvement is not consistent across evaluation targets.
Our results are mostly consistent with their recommendation to avoid sampling, but confirm their results with a stronger reference point instead of treating the full-set sparse evaluations as correct, and show that popularity-weighted sampling can provide benefit in some situations.

The MovieLens results also suggest that sparse evaluation can be useful under the appropriate design (full candidates, binary threshold), but the gap between short and long list length suggests that sparse evaluation is more favorable at larger cutoff values.
For KuaiRec, no design modifications on sparse evaluation produced positive correlation with the dense evaluation. 

Overall, these observations suggest that commonly observed differences between recommendation models may be driven as much by evaluation design as by model effectiveness, and that models identified as strongest under sparse evaluation are not necessarily those that best reflect user preferences as measured by dense evaluation.
The variation in results across dense evaluation targets further suggests that sparse offline evaluation does not consistently capture a single effectiveness construct, and that different targets may require different evaluation designs, rather than there being a uniformly best offline evaluation setup.

\section{Conclusion}
We studied to what extent sparse offline evaluation correlates with dense ground-truth evaluation, and whether modifications to evaluation design can reduce the gap.
Across two datasets and different evaluation designs, our results show that the relationship depends on the dataset and specific evaluation targets, with correlations ranging from weakly positive to strongly negative.
Sparse evaluation mostly shows positive correlation with dense evaluation for MovieLens; however, it nearly reverses the model ranking from dense evaluation for KuaiRec. 
These results suggest that the validity of sparse evaluation is not consistent across settings, and that observed differences between recommendation models may be influenced by evaluation design as much as by model effectiveness.

% \section{Appendix}
% We include the DCG-based $\tau$ correlation table in the appendix as a complement to the NDCG-based $\tau$ table for MovieLens, as unbiased DCG estimates have been shown to strongly correlate with online reward \cite{jeunenNormalisedDiscountedCumulative2024}.

% \begin{table*}[bt]
% \centering
% \caption{Kendall's $\tau$ between sparse and dense evaluation model rankings by DCG (MovieLens). Cells show $\tau$ with bootstrapped 95\% confidence intervals. \textbf{Bold} indicates highest $\tau$ per dense target column.}
% \label{tab:hero_ml_dcg}
% \resizebox{\textwidth}{!}{
% \input{tables/hero_table_ml_dcg}
% }
% \end{table*}

% \begin{table*}[bt]
% \centering
% \caption{Kendall's $\tau$ between sparse and dense evaluation model rankings by DCG (KuaiRec). Cells show $\tau$ with bootstrapped 95\% confidence intervals. \textbf{Bold} indicates highest $\tau$ per dense target column.}
% \label{tab:hero_kr_dcg}
% \resizebox{\textwidth}{!}{
% \input{tables/hero_table_kr_dcg}
% }
% \end{table*}

\begin{acks}
This work was supported by the National Science Foundation under Grant IIS 24-09199.
\end{acks}

\bibliographystyle{ACM-Reference-Format}
\bibliography{sample-base}

\end{document}

%% file: tables/models_table.tex
\begin{tabular}{llr}
\textbf{Model} & \textbf{Variants} \\
\midrule
Popular & --- \\
\midrule
Bias & --- \\
\midrule
UserKNN & $k \in \{5, 20\}$, $k_{\mathrm{min}}\in \{1, 2, 5\}$ \\
\midrule
ItemKNN (implicit) & $k \in \{5, 20\}$, $k_{\mathrm{min}}\in \{1, 2, 5\}$ \\
\midrule
ItemKNN (explicit) & $k \in \{5, 20\}$, $k_{\mathrm{min}} \in \{1, 2, 5\}$ \\
\midrule
Implicit MF (ALS) & $d \in \{32, 64\}$ \\
\midrule
Biased MF (ALS) & $d \in \{32, 64\}$ \\
\midrule
BiasedSVD & $d \in \{32, 64\}$ \\
\midrule
NMF & $d \in \{32, 64\}$ \\
\midrule
BPR\footnote{Implemented by LensKit FlexMF.} & $d \in \{32, 64, 128, 256\}$ \\
\midrule
WARP\textsuperscript{\it a} & $d \in \{32, 64, 128, 256\}$ \\
\midrule
Logistic MF\textsuperscript{\it a} & $d \in \{32, 64, 128, 256\}$ \\
\midrule
Biased MF (SGD)\textsuperscript{\it a} & $d \in \{32, 64\}$ \\
\midrule
fsSLIM & $k \in \{50, 100, 1000\}$ \\
\midrule
Total & 45 \\
\bottomrule
\end{tabular}

%% file: tables/hero_table_ml_ndcg.tex
\begin{tabular}{llrrrrrrr}
\toprule
\multicolumn{3}{c}{} & \multicolumn{3}{c}{$k=10$} & \multicolumn{3}{c}{$k=100$} \\
\cmidrule(lr){4-6}
\cmidrule(lr){7-9}
Candidate set & $k$ & Threshold & Interested & Very interested & Graded & Interested & Very interested & Graded \\
\midrule
\multirow{9}{*}{Full} & \multirow{3}{*}{10} & $\geq 3.0$ & 0.11 {\scriptsize (-0.06, 0.28)} & 0.21 {\scriptsize (0.05, 0.36)} & 0.07 {\scriptsize (-0.09, 0.24)} & 0.28 {\scriptsize (0.11, 0.43)} & 0.38 {\scriptsize (0.22, 0.52)} & 0.24 {\scriptsize (0.08, 0.40)} \\
 &  & $\geq 4.0$ & 0.09 {\scriptsize (-0.07, 0.26)} & 0.21 {\scriptsize (0.05, 0.35)} & 0.05 {\scriptsize (-0.12, 0.21)} & 0.26 {\scriptsize (0.10, 0.41)} & 0.37 {\scriptsize (0.22, 0.52)} & 0.22 {\scriptsize (0.05, 0.37)} \\
 &  & Graded & 0.10 {\scriptsize (-0.06, 0.27)} & 0.22 {\scriptsize (0.06, 0.36)} & 0.06 {\scriptsize (-0.10, 0.23)} & 0.27 {\scriptsize (0.11, 0.42)} & 0.38 {\scriptsize (0.23, 0.53)} & 0.23 {\scriptsize (0.07, 0.38)} \\
 & \multirow{3}{*}{20} & $\geq 3.0$ & 0.12 {\scriptsize (-0.05, 0.29)} & 0.23 {\scriptsize (0.06, 0.37)} & 0.08 {\scriptsize (-0.08, 0.24)} & 0.28 {\scriptsize (0.11, 0.44)} & 0.39 {\scriptsize (0.23, 0.54)} & \textbf{0.24 {\scriptsize (0.07, 0.40)}} \\
 &  & $\geq 4.0$ & 0.10 {\scriptsize (-0.06, 0.26)} & 0.22 {\scriptsize (0.07, 0.36)} & 0.05 {\scriptsize (-0.11, 0.21)} & 0.26 {\scriptsize (0.10, 0.41)} & 0.38 {\scriptsize (0.23, 0.52)} & 0.22 {\scriptsize (0.06, 0.37)} \\
 &  & Graded & 0.10 {\scriptsize (-0.06, 0.27)} & 0.23 {\scriptsize (0.07, 0.36)} & 0.06 {\scriptsize (-0.11, 0.23)} & 0.27 {\scriptsize (0.11, 0.42)} & 0.39 {\scriptsize (0.23, 0.52)} & 0.23 {\scriptsize (0.07, 0.38)} \\
 & \multirow{3}{*}{100} & $\geq 3.0$ & \textbf{0.12 {\scriptsize (-0.04, 0.28)}} & 0.26 {\scriptsize (0.10, 0.40)} & \textbf{0.08 {\scriptsize (-0.08, 0.25)}} & \textbf{0.29 {\scriptsize (0.12, 0.44)}} & \textbf{0.43 {\scriptsize (0.26, 0.57)}} & 0.24 {\scriptsize (0.07, 0.39)} \\
 &  & $\geq 4.0$ & 0.08 {\scriptsize (-0.08, 0.24)} & 0.22 {\scriptsize (0.07, 0.36)} & 0.03 {\scriptsize (-0.13, 0.19)} & 0.24 {\scriptsize (0.08, 0.39)} & 0.38 {\scriptsize (0.22, 0.53)} & 0.19 {\scriptsize (0.03, 0.34)} \\
 &  & Graded & 0.09 {\scriptsize (-0.07, 0.26)} & 0.23 {\scriptsize (0.08, 0.37)} & 0.04 {\scriptsize (-0.11, 0.21)} & 0.25 {\scriptsize (0.09, 0.40)} & 0.40 {\scriptsize (0.23, 0.54)} & 0.20 {\scriptsize (0.04, 0.36)} \\
\midrule
\multirow{9}{*}{U-1000} & \multirow{3}{*}{10} & $\geq 3.0$ & 0.07 {\scriptsize (-0.08, 0.23)} & 0.24 {\scriptsize (0.08, 0.37)} & 0.01 {\scriptsize (-0.14, 0.17)} & 0.18 {\scriptsize (0.01, 0.34)} & 0.37 {\scriptsize (0.21, 0.52)} & 0.13 {\scriptsize (-0.03, 0.29)} \\
 &  & $\geq 4.0$ & 0.02 {\scriptsize (-0.13, 0.17)} & 0.19 {\scriptsize (0.04, 0.33)} & -0.04 {\scriptsize (-0.19, 0.11)} & 0.14 {\scriptsize (-0.01, 0.30)} & 0.33 {\scriptsize (0.17, 0.46)} & 0.09 {\scriptsize (-0.06, 0.24)} \\
 &  & Graded & 0.03 {\scriptsize (-0.11, 0.19)} & 0.20 {\scriptsize (0.06, 0.34)} & -0.02 {\scriptsize (-0.17, 0.14)} & 0.15 {\scriptsize (0.00, 0.31)} & 0.34 {\scriptsize (0.19, 0.48)} & 0.11 {\scriptsize (-0.05, 0.26)} \\
 & \multirow{3}{*}{20} & $\geq 3.0$ & 0.07 {\scriptsize (-0.08, 0.22)} & 0.23 {\scriptsize (0.08, 0.37)} & 0.01 {\scriptsize (-0.14, 0.17)} & 0.17 {\scriptsize (0.00, 0.33)} & 0.37 {\scriptsize (0.21, 0.51)} & 0.12 {\scriptsize (-0.04, 0.28)} \\
 &  & $\geq 4.0$ & 0.00 {\scriptsize (-0.14, 0.15)} & 0.18 {\scriptsize (0.03, 0.32)} & -0.05 {\scriptsize (-0.20, 0.10)} & 0.12 {\scriptsize (-0.04, 0.27)} & 0.31 {\scriptsize (0.16, 0.45)} & 0.07 {\scriptsize (-0.09, 0.22)} \\
 &  & Graded & 0.02 {\scriptsize (-0.13, 0.17)} & 0.20 {\scriptsize (0.05, 0.33)} & -0.04 {\scriptsize (-0.18, 0.12)} & 0.13 {\scriptsize (-0.03, 0.29)} & 0.33 {\scriptsize (0.17, 0.47)} & 0.08 {\scriptsize (-0.07, 0.24)} \\
 & \multirow{3}{*}{100} & $\geq 3.0$ & 0.06 {\scriptsize (-0.09, 0.20)} & 0.23 {\scriptsize (0.08, 0.36)} & 0.01 {\scriptsize (-0.14, 0.16)} & 0.15 {\scriptsize (-0.02, 0.31)} & 0.36 {\scriptsize (0.20, 0.50)} & 0.11 {\scriptsize (-0.05, 0.26)} \\
 &  & $\geq 4.0$ & -0.02 {\scriptsize (-0.16, 0.13)} & 0.17 {\scriptsize (0.02, 0.30)} & -0.07 {\scriptsize (-0.21, 0.08)} & 0.08 {\scriptsize (-0.08, 0.23)} & 0.29 {\scriptsize (0.13, 0.43)} & 0.03 {\scriptsize (-0.12, 0.19)} \\
 &  & Graded & 0.00 {\scriptsize (-0.14, 0.15)} & 0.19 {\scriptsize (0.04, 0.32)} & -0.05 {\scriptsize (-0.20, 0.10)} & 0.10 {\scriptsize (-0.06, 0.26)} & 0.31 {\scriptsize (0.15, 0.45)} & 0.05 {\scriptsize (-0.11, 0.20)} \\
\midrule
\multirow{9}{*}{P-1000} & \multirow{3}{*}{10} & $\geq 3.0$ & 0.06 {\scriptsize (-0.09, 0.21)} & 0.28 {\scriptsize (0.13, 0.41)} & 0.02 {\scriptsize (-0.13, 0.17)} & -0.02 {\scriptsize (-0.17, 0.15)} & 0.20 {\scriptsize (0.03, 0.35)} & -0.05 {\scriptsize (-0.20, 0.12)} \\
 &  & $\geq 4.0$ & 0.07 {\scriptsize (-0.08, 0.19)} & 0.27 {\scriptsize (0.14, 0.38)} & 0.05 {\scriptsize (-0.09, 0.18)} & 0.05 {\scriptsize (-0.11, 0.19)} & 0.26 {\scriptsize (0.10, 0.38)} & 0.01 {\scriptsize (-0.14, 0.16)} \\
 &  & Graded & 0.07 {\scriptsize (-0.07, 0.21)} & 0.28 {\scriptsize (0.15, 0.39)} & 0.06 {\scriptsize (-0.08, 0.20)} & 0.04 {\scriptsize (-0.11, 0.19)} & 0.25 {\scriptsize (0.09, 0.39)} & 0.01 {\scriptsize (-0.14, 0.16)} \\
 & \multirow{3}{*}{20} & $\geq 3.0$ & 0.06 {\scriptsize (-0.09, 0.21)} & 0.29 {\scriptsize (0.13, 0.42)} & 0.02 {\scriptsize (-0.12, 0.17)} & -0.03 {\scriptsize (-0.18, 0.13)} & 0.19 {\scriptsize (0.03, 0.35)} & -0.06 {\scriptsize (-0.21, 0.10)} \\
 &  & $\geq 4.0$ & 0.06 {\scriptsize (-0.08, 0.19)} & 0.27 {\scriptsize (0.14, 0.37)} & 0.04 {\scriptsize (-0.09, 0.18)} & 0.03 {\scriptsize (-0.13, 0.17)} & 0.24 {\scriptsize (0.09, 0.38)} & -0.01 {\scriptsize (-0.16, 0.13)} \\
 &  & Graded & 0.07 {\scriptsize (-0.07, 0.20)} & 0.28 {\scriptsize (0.15, 0.39)} & 0.06 {\scriptsize (-0.09, 0.19)} & 0.02 {\scriptsize (-0.13, 0.16)} & 0.24 {\scriptsize (0.08, 0.37)} & -0.01 {\scriptsize (-0.16, 0.13)} \\
 & \multirow{3}{*}{100} & $\geq 3.0$ & 0.05 {\scriptsize (-0.10, 0.20)} & \textbf{0.29 {\scriptsize (0.14, 0.41)}} & 0.01 {\scriptsize (-0.13, 0.16)} & -0.05 {\scriptsize (-0.20, 0.10)} & 0.18 {\scriptsize (0.01, 0.33)} & -0.08 {\scriptsize (-0.22, 0.07)} \\
 &  & $\geq 4.0$ & 0.05 {\scriptsize (-0.09, 0.18)} & 0.27 {\scriptsize (0.13, 0.37)} & 0.04 {\scriptsize (-0.10, 0.17)} & -0.00 {\scriptsize (-0.14, 0.14)} & 0.22 {\scriptsize (0.07, 0.36)} & -0.03 {\scriptsize (-0.18, 0.10)} \\
 &  & Graded & 0.05 {\scriptsize (-0.09, 0.19)} & 0.27 {\scriptsize (0.13, 0.38)} & 0.04 {\scriptsize (-0.10, 0.18)} & -0.02 {\scriptsize (-0.16, 0.13)} & 0.21 {\scriptsize (0.05, 0.34)} & -0.05 {\scriptsize (-0.19, 0.09)} \\
\bottomrule
\end{tabular}

%% file: tables/best_dense_rank_ml.tex
\begin{tabular}{lrrrrr}
\toprule
Model & Dense NDCG@10 & Sparse NDCG (P-1000, $k=100$, $r\geq3.0$) & Rank (best corr.) & Best rank (any) & Best design \\
\midrule
\textbf{BiasedSVD-d64} & \textbf{0.323} & \textbf{0.177} & \textbf{40} & \textbf{23} & Full, $k=20$, $r\geq4.0$ \\
BiasedSVD-d32 & 0.277 & 0.179 & 38 & 22 & Full, $k=10$, Graded \\
fsSLIM-nn1000 & 0.242 & 0.267 & 1 & 1 & U-1000, $k=10$, $r\geq3.0$ \\
fsSLIM-nn50 & 0.231 & 0.263 & 3 & 3 & U-1000, $k=10$, $r\geq3.0$ \\
fsSLIM-nn100 & 0.231 & 0.265 & 2 & 1 & U-1000, $k=100$, $r\geq3.0$ \\
\bottomrule
\end{tabular}

%% file: tables/hero_table_kr_ndcg.tex
\begin{tabular}{llrrrrrrr}
\toprule
\multicolumn{3}{c}{} & \multicolumn{3}{c}{$k=10$} & \multicolumn{3}{c}{$k=100$} \\
\cmidrule(lr){4-6}
\cmidrule(lr){7-9}
Candidate set & $k$ & Threshold & Watched $\geq$85\% & Watched 2$\times$ & Graded & Watched $\geq$85\% & Watched 2$\times$ & Graded \\
\midrule
\multirow{9}{*}{Full} & \multirow{3}{*}{10} & $\geq 0.85$ & -0.36 {\scriptsize (-0.45, -0.25)} & -0.45 {\scriptsize (-0.54, -0.35)} & -0.48 {\scriptsize (-0.56, -0.36)} & -0.38 {\scriptsize (-0.48, -0.27)} & -0.49 {\scriptsize (-0.57, -0.38)} & -0.50 {\scriptsize (-0.58, -0.38)} \\
 &  & $\geq 2.0$ & -0.18 {\scriptsize (-0.29, -0.05)} & -0.25 {\scriptsize (-0.36, -0.13)} & -0.28 {\scriptsize (-0.39, -0.15)} & -0.24 {\scriptsize (-0.35, -0.11)} & -0.28 {\scriptsize (-0.39, -0.15)} & -0.31 {\scriptsize (-0.43, -0.19)} \\
 &  & Graded & -0.36 {\scriptsize (-0.45, -0.25)} & -0.45 {\scriptsize (-0.54, -0.35)} & -0.48 {\scriptsize (-0.57, -0.37)} & -0.38 {\scriptsize (-0.48, -0.27)} & -0.49 {\scriptsize (-0.57, -0.37)} & -0.50 {\scriptsize (-0.58, -0.38)} \\
 & \multirow{3}{*}{20} & $\geq 0.85$ & -0.37 {\scriptsize (-0.46, -0.27)} & -0.47 {\scriptsize (-0.55, -0.36)} & -0.49 {\scriptsize (-0.57, -0.39)} & -0.39 {\scriptsize (-0.48, -0.28)} & -0.50 {\scriptsize (-0.59, -0.39)} & -0.51 {\scriptsize (-0.59, -0.40)} \\
 &  & $\geq 2.0$ & \textbf{-0.17 {\scriptsize (-0.29, -0.05)}} & -0.24 {\scriptsize (-0.35, -0.11)} & -0.27 {\scriptsize (-0.38, -0.14)} & \textbf{-0.24 {\scriptsize (-0.35, -0.11)}} & -0.27 {\scriptsize (-0.38, -0.14)} & -0.31 {\scriptsize (-0.42, -0.18)} \\
 &  & Graded & -0.37 {\scriptsize (-0.46, -0.27)} & -0.47 {\scriptsize (-0.55, -0.36)} & -0.49 {\scriptsize (-0.57, -0.39)} & -0.40 {\scriptsize (-0.48, -0.29)} & -0.50 {\scriptsize (-0.58, -0.39)} & -0.51 {\scriptsize (-0.59, -0.40)} \\
 & \multirow{3}{*}{100} & $\geq 0.85$ & -0.42 {\scriptsize (-0.50, -0.34)} & -0.52 {\scriptsize (-0.58, -0.43)} & -0.54 {\scriptsize (-0.61, -0.46)} & -0.44 {\scriptsize (-0.51, -0.36)} & -0.55 {\scriptsize (-0.62, -0.46)} & -0.56 {\scriptsize (-0.62, -0.47)} \\
 &  & $\geq 2.0$ & -0.21 {\scriptsize (-0.31, -0.10)} & -0.24 {\scriptsize (-0.34, -0.12)} & -0.27 {\scriptsize (-0.37, -0.15)} & -0.27 {\scriptsize (-0.38, -0.16)} & -0.26 {\scriptsize (-0.37, -0.14)} & -0.30 {\scriptsize (-0.41, -0.19)} \\
 &  & Graded & -0.43 {\scriptsize (-0.51, -0.35)} & -0.53 {\scriptsize (-0.59, -0.45)} & -0.55 {\scriptsize (-0.62, -0.47)} & -0.45 {\scriptsize (-0.52, -0.38)} & -0.56 {\scriptsize (-0.63, -0.47)} & -0.57 {\scriptsize (-0.63, -0.49)} \\
\midrule
\multirow{9}{*}{U-1000} & \multirow{3}{*}{10} & $\geq 0.85$ & -0.42 {\scriptsize (-0.49, -0.34)} & -0.51 {\scriptsize (-0.58, -0.43)} & -0.54 {\scriptsize (-0.60, -0.45)} & -0.43 {\scriptsize (-0.50, -0.35)} & -0.55 {\scriptsize (-0.62, -0.46)} & -0.55 {\scriptsize (-0.61, -0.47)} \\
 &  & $\geq 2.0$ & -0.19 {\scriptsize (-0.29, -0.08)} & -0.19 {\scriptsize (-0.30, -0.09)} & -0.22 {\scriptsize (-0.33, -0.11)} & -0.25 {\scriptsize (-0.35, -0.14)} & -0.21 {\scriptsize (-0.32, -0.10)} & -0.26 {\scriptsize (-0.36, -0.14)} \\
 &  & Graded & -0.45 {\scriptsize (-0.52, -0.37)} & -0.54 {\scriptsize (-0.60, -0.46)} & -0.56 {\scriptsize (-0.63, -0.48)} & -0.46 {\scriptsize (-0.53, -0.39)} & -0.57 {\scriptsize (-0.64, -0.49)} & -0.58 {\scriptsize (-0.64, -0.50)} \\
 & \multirow{3}{*}{20} & $\geq 0.85$ & -0.43 {\scriptsize (-0.51, -0.35)} & -0.52 {\scriptsize (-0.59, -0.44)} & -0.55 {\scriptsize (-0.62, -0.46)} & -0.44 {\scriptsize (-0.51, -0.37)} & -0.55 {\scriptsize (-0.62, -0.47)} & -0.56 {\scriptsize (-0.62, -0.48)} \\
 &  & $\geq 2.0$ & -0.21 {\scriptsize (-0.31, -0.11)} & -0.21 {\scriptsize (-0.31, -0.11)} & -0.25 {\scriptsize (-0.35, -0.14)} & -0.27 {\scriptsize (-0.37, -0.17)} & -0.23 {\scriptsize (-0.33, -0.13)} & -0.28 {\scriptsize (-0.38, -0.17)} \\
 &  & Graded & -0.46 {\scriptsize (-0.53, -0.39)} & -0.55 {\scriptsize (-0.61, -0.47)} & -0.58 {\scriptsize (-0.64, -0.50)} & -0.48 {\scriptsize (-0.54, -0.41)} & -0.59 {\scriptsize (-0.65, -0.50)} & -0.59 {\scriptsize (-0.65, -0.52)} \\
 & \multirow{3}{*}{100} & $\geq 0.85$ & -0.41 {\scriptsize (-0.49, -0.34)} & -0.50 {\scriptsize (-0.57, -0.42)} & -0.53 {\scriptsize (-0.60, -0.45)} & -0.42 {\scriptsize (-0.49, -0.35)} & -0.54 {\scriptsize (-0.61, -0.45)} & -0.54 {\scriptsize (-0.60, -0.46)} \\
 &  & $\geq 2.0$ & -0.23 {\scriptsize (-0.33, -0.13)} & -0.24 {\scriptsize (-0.33, -0.13)} & -0.27 {\scriptsize (-0.36, -0.17)} & -0.29 {\scriptsize (-0.38, -0.18)} & -0.26 {\scriptsize (-0.35, -0.15)} & -0.30 {\scriptsize (-0.40, -0.19)} \\
 &  & Graded & -0.46 {\scriptsize (-0.52, -0.38)} & -0.54 {\scriptsize (-0.60, -0.46)} & -0.57 {\scriptsize (-0.63, -0.50)} & -0.47 {\scriptsize (-0.53, -0.40)} & -0.58 {\scriptsize (-0.64, -0.50)} & -0.59 {\scriptsize (-0.64, -0.51)} \\
\midrule
\multirow{9}{*}{P-1000} & \multirow{3}{*}{10} & $\geq 0.85$ & -0.43 {\scriptsize (-0.51, -0.34)} & -0.50 {\scriptsize (-0.57, -0.41)} & -0.53 {\scriptsize (-0.60, -0.45)} & -0.45 {\scriptsize (-0.52, -0.37)} & -0.53 {\scriptsize (-0.60, -0.44)} & -0.55 {\scriptsize (-0.62, -0.47)} \\
 &  & $\geq 2.0$ & -0.20 {\scriptsize (-0.29, -0.09)} & -0.19 {\scriptsize (-0.28, -0.09)} & -0.22 {\scriptsize (-0.31, -0.12)} & -0.25 {\scriptsize (-0.33, -0.15)} & -0.21 {\scriptsize (-0.30, -0.10)} & -0.25 {\scriptsize (-0.34, -0.15)} \\
 &  & Graded & -0.45 {\scriptsize (-0.53, -0.37)} & -0.51 {\scriptsize (-0.58, -0.43)} & -0.55 {\scriptsize (-0.62, -0.46)} & -0.48 {\scriptsize (-0.54, -0.40)} & -0.54 {\scriptsize (-0.61, -0.46)} & -0.57 {\scriptsize (-0.63, -0.49)} \\
 & \multirow{3}{*}{20} & $\geq 0.85$ & -0.43 {\scriptsize (-0.51, -0.35)} & -0.49 {\scriptsize (-0.56, -0.41)} & -0.52 {\scriptsize (-0.59, -0.44)} & -0.45 {\scriptsize (-0.52, -0.37)} & -0.52 {\scriptsize (-0.59, -0.44)} & -0.54 {\scriptsize (-0.61, -0.46)} \\
 &  & $\geq 2.0$ & -0.20 {\scriptsize (-0.29, -0.10)} & \textbf{-0.19 {\scriptsize (-0.28, -0.09)}} & \textbf{-0.22 {\scriptsize (-0.31, -0.12)}} & -0.25 {\scriptsize (-0.34, -0.15)} & \textbf{-0.21 {\scriptsize (-0.29, -0.11)}} & \textbf{-0.25 {\scriptsize (-0.34, -0.15)}} \\
 &  & Graded & -0.46 {\scriptsize (-0.54, -0.38)} & -0.51 {\scriptsize (-0.57, -0.43)} & -0.54 {\scriptsize (-0.61, -0.47)} & -0.49 {\scriptsize (-0.55, -0.41)} & -0.54 {\scriptsize (-0.60, -0.46)} & -0.57 {\scriptsize (-0.63, -0.50)} \\
 & \multirow{3}{*}{100} & $\geq 0.85$ & -0.42 {\scriptsize (-0.51, -0.33)} & -0.45 {\scriptsize (-0.52, -0.36)} & -0.49 {\scriptsize (-0.56, -0.40)} & -0.45 {\scriptsize (-0.53, -0.36)} & -0.47 {\scriptsize (-0.54, -0.38)} & -0.52 {\scriptsize (-0.59, -0.43)} \\
 &  & $\geq 2.0$ & -0.22 {\scriptsize (-0.31, -0.12)} & -0.21 {\scriptsize (-0.30, -0.11)} & -0.25 {\scriptsize (-0.34, -0.14)} & -0.27 {\scriptsize (-0.36, -0.16)} & -0.23 {\scriptsize (-0.32, -0.13)} & -0.28 {\scriptsize (-0.36, -0.17)} \\
 &  & Graded & -0.46 {\scriptsize (-0.54, -0.37)} & -0.47 {\scriptsize (-0.54, -0.39)} & -0.51 {\scriptsize (-0.58, -0.43)} & -0.49 {\scriptsize (-0.56, -0.41)} & -0.50 {\scriptsize (-0.56, -0.42)} & -0.54 {\scriptsize (-0.61, -0.46)} \\
\bottomrule
\end{tabular}

%% file: tables/best_dense_rank_kr.tex
\begin{tabular}{lrrrrr}
\toprule
Model & Dense NDCG@10 & Sparse NDCG (Full, $k=20$, $r\geq2.0$) & Rank (best corr.) & Best rank (any) & Best design \\
\midrule
\textbf{BiasedSVD-d32} & \textbf{0.672} & \textbf{0.0} & \textbf{35} & \textbf{27} & P-1000, $k=100$, $r\geq2.0$ \\
BiasedSVD-d64 & 0.665 & 0.001 & 32 & 26 & P-1000, $k=100$, $r\geq2.0$ \\
Bias & 0.654 & 0.0 & 44 & 28 & P-1000, $k=10$, $r\geq2.0$ \\
ItemKNN-nnbrs5-min1-imp & 0.649 & 0.059 & 18 & 18 & Full, $k=20$, $r\geq2.0$ \\
ItemKNN-nnbrs5-min2-imp & 0.648 & 0.059 & 17 & 17 & Full, $k=20$, $r\geq2.0$ \\
\bottomrule
\end{tabular}

%% file: sample-base.bib
@article{carraroSamplingApproachDebiasing2022,
  title = {A Sampling Approach to {{Debiasing}} the Offline Evaluation of Recommender Systems},
  author = {Carraro, Diego and Bridge, Derek},
  year = 2022,
  month = apr,
  journal = {Journal of Intelligent Information Systems},
  volume = {58},
  number = {2},
  pages = {311--336},
  issn = {0925-9902, 1573-7675},
  doi = {10.1007/s10844-021-00651-y},
  urldate = {2026-04-02},
  langid = {english}
}

@inproceedings{gaoKuaiRecFullyobservedDataset2022,
  title = {{{KuaiRec}}: {{A Fully-observed Dataset}} and {{Insights}} for {{Evaluating Recommender Systems}}},
  shorttitle = {{{KuaiRec}}},
  booktitle = {Proceedings of the 31st {{ACM International Conference}} on {{Information}} \& {{Knowledge Management}}},
  author = {Gao, Chongming and Li, Shijun and Lei, Wenqiang and Chen, Jiawei and Li, Biao and Jiang, Peng and He, Xiangnan and Mao, Jiaxin and Chua, Tat-Seng},
  year = 2022,
  month = oct,
  pages = {540--550},
  publisher = {ACM},
  address = {Atlanta, GA, USA},
  doi = {10.1145/3511808.3557220},
  urldate = {2026-04-06},
  isbn = {978-1-4503-9236-5},
  langid = {english}
}

@inproceedings{gusakTimeSplitExploring2025,
  title = {Time to {{Split}}: {{Exploring Data Splitting Strategies}} for {{Offline Evaluation}} of {{Sequential Recommenders}}},
  shorttitle = {Time to {{Split}}},
  booktitle = {Proceedings of the {{Nineteenth ACM Conference}} on {{Recommender Systems}}},
  author = {Gusak, Danil and Volodkevich, Anna and Klenitskiy, Anton and Vasilev, Alexey and Frolov, Evgeny},
  year = 2025,
  month = sep,
  pages = {874--883},
  publisher = {ACM},
  address = {Prague, Czech Republic},
  doi = {10.1145/3705328.3748164},
  urldate = {2026-02-24},
  isbn = {979-8-4007-1364-4},
  langid = {english}
}

@inproceedings{mengExploringDataSplitting2020,
  title = {Exploring {{Data Splitting Strategies}} for the {{Evaluation}} of {{Recommendation Models}}},
  booktitle = {Fourteenth {{ACM Conference}} on {{Recommender Systems}}},
  author = {Meng, Zaiqiao and McCreadie, Richard and Macdonald, Craig and Ounis, Iadh},
  year = 2020,
  month = sep,
  pages = {681--686},
  publisher = {ACM},
  address = {Virtual Event, Brazil},
  doi = {10.1145/3383313.3418479},
  urldate = {2026-02-24},
  copyright = {https://www.acm.org/publications/policies/copyright\_policy\#Background},
  isbn = {978-1-4503-7583-2},
  langid = {english}
}

@inproceedings{smuckerExtendingMovieLens32MProvide2025,
  title = {Extending {{MovieLens-32M}} to {{Provide New Evaluation Objectives}}},
  booktitle = {Proceedings of the 48th {{International ACM SIGIR Conference}} on {{Research}} and {{Development}} in {{Information Retrieval}}},
  author = {Smucker, Mark D. and Chamani, Houmaan},
  year = 2025,
  month = jul,
  pages = {3520--3529},
  publisher = {ACM},
  address = {Padua, Italy},
  doi = {10.1145/3726302.3730328},
  urldate = {2026-04-06},
  isbn = {979-8-4007-1592-1},
  langid = {english}
}

@inproceedings{steckTrainingTestingRecommender2010,
  title = {Training and Testing of Recommender Systems on Data Missing Not at Random},
  booktitle = {Proceedings of the 16th {{ACM SIGKDD}} International Conference on {{Knowledge}} Discovery and Data Mining},
  author = {Steck, Harald},
  year = 2010,
  month = jul,
  pages = {713--722},
  publisher = {ACM},
  address = {Washington DC USA},
  doi = {10.1145/1835804.1835895},
  urldate = {2026-04-02},
  isbn = {978-1-4503-0055-1},
  langid = {english}
}

@article{belloginStatisticalBiasesInformation2017,
  title = {Statistical Biases in {{Information Retrieval}} Metrics for Recommender Systems},
  author = {Bellog{\'i}n, Alejandro and Castells, Pablo and Cantador, Iv{\'a}n},
  year = 2017,
  month = dec,
  journal = {Information Retrieval Journal},
  volume = {20},
  number = {6},
  pages = {606--634},
  issn = {1386-4564, 1573-7659},
  doi = {10.1007/s10791-017-9312-z},
  urldate = {2026-04-07},
  langid = {english}
}

@inproceedings{ekstrand2020lenskit,
  title={{{LensKit}} for {{Python}}: Next-generation software for recommender systems experiments},
  author={Ekstrand, Michael D},
  booktitle={Proceedings of the 29th ACM international conference on information \& knowledge management},
  pages={2999--3006},
  year={2020}
}

@inproceedings{canamaresTargetItemSampling2020,
  title = {On {{Target Item Sampling}} in {{Offline Recommender System Evaluation}}},
  booktitle = {{{RecSys}} '20},
  author = {Ca{\~n}amares, Roc{\'i}o and Castells, Pablo},
  year = 2020,
  month = sep,
  pages = {259--268},
  publisher = {ACM},
  address = {New York, NY, USA},
  doi = {10.1145/3383313.3412259},
  urldate = {2020-09-23}
}

@inproceedings{ekstrandSturgeonCoolKids2017,
  title = {Sturgeon and the {{Cool Kids}}: {{Problems}} with {{Top-N Recommender Evaluation}}},
  booktitle = {Proceedings of the 30th {{Florida Artificial Intelligence Research Society Conference}}},
  author = {Ekstrand, Michael D and Mahant, Vaibhav},
  year = 2017,
  month = may,
  series = {{{FLAIRS}} 30},
  publisher = {AAAI Press}
}

@inproceedings{ihemelanduCandidateSetSampling2023,
  title = {Candidate Set Sampling for Evaluating Top-{{N}} Recommendation},
  booktitle = {Proceedings of the 22nd {{IEEE}}/{{WIC}} International Conference on Web Intelligence and Intelligent Agent Technology},
  author = {Ihemelandu, Ngozi and Ekstrand, Michael D.},
  year = 2023,
  month = oct,
  eprint = {2309.11723},
  primaryclass = {cs.IR},
  pages = {88--94},
  doi = {10.1109/WI-IAT59888.2023.00018},
  urldate = {2023-11-08},
  archiveprefix = {arXiv}
}

@incollection{kricheneSampledMetricsItem2020,
  title = {On {{Sampled Metrics}} for {{Item Recommendation}}},
  booktitle = {Proceedings of the 26th {{ACM SIGKDD International Conference}} on {{Knowledge Discovery}} \& {{Data Mining}}},
  author = {Krichene, Walid and Rendle, Steffen},
  year = 2020,
  month = aug,
  pages = {1748--1757},
  publisher = {ACM},
  address = {New York, NY, USA},
  doi = {10.1145/3394486.3403226},
  urldate = {2021-07-27},
  isbn = {978-1-4503-7998-4}
}

@inproceedings{pereiraReliabilitySamplingStrategies2025,
author = {Pereira, Bruno L. and Said, Alan and Santos, Rodrygo L. T.},
title = {On the Reliability of Sampling Strategies in Offline Recommender Evaluation},
year = 2025,
isbn = {9798400713644},
publisher = {ACM},
address = {New York, NY, USA},
url = {https://doi.org/10.1145/3705328.3748086},
doi = {10.1145/3705328.3748086},
booktitle = {Proceedings of the Nineteenth ACM Conference on Recommender Systems},
pages = {360–369},
numpages = {10},
location = {
},
series = {RecSys '25}
}

@inproceedings{rossettiContrastingOfflineOnline2016,
  title = {Contrasting {{Offline}} and {{Online Results}} When {{Evaluating Recommendation Algorithms}}},
  booktitle = {Proceedings of the 10th {{ACM Conference}} on {{Recommender Systems}}},
  author = {Rossetti, Marco and Stella, Fabio and Zanker, Markus},
  year = 2016,
  month = sep,
  pages = {31--34},
  publisher = {ACM},
  address = {Boston, Massachusetts, USA},
  doi = {10.1145/2959100.2959176},
  urldate = {2026-04-10},
  isbn = {978-1-4503-4035-9},
  langid = {english}
}

@inproceedings{sunTakeFreshLook2023,
  title = {Take a {{Fresh Look}} at {{Recommender Systems}} from an {{Evaluation Standpoint}}},
  booktitle = {Proceedings of the 46th {{International ACM SIGIR Conference}} on {{Research}} and {{Development}} in {{Information Retrieval}}},
  author = {Sun, Aixin},
  year = 2023,
  month = jul,
  pages = {2629--2638},
  publisher = {ACM},
  address = {Taipei Taiwan},
  doi = {10.1145/3539618.3591931},
  urldate = {2025-08-19},
  isbn = {978-1-4503-9408-6},
  langid = {english}
}

@inproceedings{verachtertrobinAreWeForgetting2022,
  title = {Are We Forgetting Something? {{Correctly}} Evaluate a Recommender System with an Optimal Training Window},
  shorttitle = {{{PERSPECTIVES}} 2022},
  booktitle = {Proceedings of the {{Perspectives}} on the {{Evaluation}} of {{Recommender Systems Workshop}} 2022},
  author = {Verachtert, Robin and Michiels, Lien and Goethals, Bart},
  year = 2022
}

@inproceedings{hidasiWidespreadFlawsOffline2023,
  title = {Widespread {{Flaws}} in {{Offline Evaluation}} of {{Recommender Systems}}},
  booktitle = {Proceedings of the 17th {{ACM Conference}} on {{Recommender Systems}}},
  author = {Hidasi, Bal{\'a}zs and Czapp, {\'A}d{\'a}m Tibor},
  year = 2023,
  month = sep,
  pages = {848--855},
  publisher = {ACM},
  address = {Singapore, Singapore},
  doi = {10.1145/3604915.3608839},
  urldate = {2026-04-14},
  isbn = {979-8-4007-0241-9},
  langid = {english}
}

@article{jadidinejadSimpsonsParadoxOffline2022,
  title = {The {{Simpson}}'s {{Paradox}} in the {{Offline Evaluation}} of {{Recommendation Systems}}},
  author = {Jadidinejad, Amir H. and Macdonald, Craig and Ounis, Iadh},
  year = 2022,
  month = jan,
  journal = {ACM Transactions on Information Systems},
  volume = {40},
  number = {1},
  pages = {1--22},
  issn = {1046-8188, 1558-2868},
  doi = {10.1145/3458509},
  urldate = {2026-04-14},
  langid = {english}
}

@inproceedings{jeunenRevisitingOfflineEvaluation2019,
  title = {Revisiting Offline Evaluation for Implicit-Feedback Recommender Systems},
  booktitle = {Proceedings of the 13th {{ACM Conference}} on {{Recommender Systems}}},
  author = {Jeunen, Olivier},
  year = 2019,
  month = sep,
  pages = {596--600},
  publisher = {ACM},
  address = {Copenhagen, Denmark},
  doi = {10.1145/3298689.3347069},
  urldate = {2026-04-14},
  isbn = {978-1-4503-6243-6},
  langid = {english}
}

@incollection{beelComparisonOfflineEvaluations2015,
  title = {A {{Comparison}} of {{Offline Evaluations}}, {{Online Evaluations}}, and {{User Studies}} in the {{Context}} of {{Research-Paper Recommender Systems}}},
  booktitle = {Research and {{Advanced Technology}} for {{Digital Libraries}}},
  author = {Beel, Joeran and Langer, Stefan},
  editor = {Kapidakis, Sarantos and Mazurek, Cezary and Werla, Marcin},
  year = 2015,
  volume = {9316},
  pages = {153--168},
  publisher = {Springer International Publishing},
  address = {Cham},
  doi = {10.1007/978-3-319-24592-8_12},
  urldate = {2026-03-26},
  isbn = {978-3-319-24591-1 978-3-319-24592-8}
}

@inproceedings{marlinCollaborativePredictionRanking2009a,
  title = {Collaborative Prediction and Ranking with Non-Random Missing Data},
  booktitle = {Proceedings of the Third {{ACM}} Conference on {{Recommender}} Systems},
  author = {Marlin, Benjamin M. and Zemel, Richard S.},
  year = 2009,
  month = oct,
  pages = {5--12},
  publisher = {ACM},
  address = {New York, New York, USA},
  doi = {10.1145/1639714.1639717},
  urldate = {2026-04-14},
  isbn = {978-1-60558-435-5},
  langid = {english}
}

@article{jiCriticalStudyData2023,
  title = {A {{Critical Study}} on {{Data Leakage}} in {{Recommender System Offline Evaluation}}},
  author = {Ji, Yitong and Sun, Aixin and Zhang, Jie and Li, Chenliang},
  year = 2023,
  month = jul,
  journal = {ACM Transactions on Information Systems},
  volume = {41},
  number = {3},
  pages = {1--27},
  issn = {1046-8188, 1558-2868},
  doi = {10.1145/3569930},
  urldate = {2026-04-14},
  langid = {english}
}

@inproceedings{korenFactorizationMeetsNeighborhood2008,
  title = {Factorization Meets the Neighborhood: A Multifaceted Collaborative Filtering Model},
  shorttitle = {Factorization Meets the Neighborhood},
  booktitle = {Proceedings of the 14th {{ACM SIGKDD}} International Conference on {{Knowledge}} Discovery and Data Mining},
  author = {Koren, Yehuda},
  year = 2008,
  month = aug,
  pages = {426--434},
  publisher = {ACM},
  address = {Las Vegas, Nevada, USA},
  doi = {10.1145/1401890.1401944},
  urldate = {2026-04-16},
  isbn = {978-1-60558-193-4},
  langid = {english}
}

@misc{marlinCollaborativeFilteringMissing2012,
  title = {Collaborative {{Filtering}} and the {{Missing}} at {{Random Assumption}}},
  author = {Marlin, Benjamin and Zemel, Richard S. and Roweis, Sam and Slaney, Malcolm},
  year = 2012,
  publisher = {arXiv},
  doi = {10.48550/arXiv.1206.5267},
  urldate = {2026-04-17},
  copyright = {arXiv.org perpetual, non-exclusive license}
}

@inproceedings{cremonesiPerformanceRecommenderAlgorithms2010,
  title = {Performance of Recommender Algorithms on Top-{{N}} Recommendation Tasks},
  booktitle = {Proceedings of the Fourth {{ACM}} Conference on {{Recommender}} Systems},
  author = {Cremonesi, Paolo and Koren, Yehuda and Turrin, Roberto},
  year = 2010,
  month = sep,
  pages = {39--46},
  publisher = {ACM},
  address = {Barcelona, Spain},
  doi = {10.1145/1864708.1864721},
  urldate = {2026-04-17},
  isbn = {978-1-60558-906-0},
  langid = {english}
}

@article{cookCurrentConceptsValidity2006,
  title = {Current {{Concepts}} in {{Validity}} and {{Reliability}} for {{Psychometric Instruments}}: {{Theory}} and {{Application}}},
  shorttitle = {Current {{Concepts}} in {{Validity}} and {{Reliability}} for {{Psychometric Instruments}}},
  author = {Cook, David A. and Beckman, Thomas J.},
  year = 2006,
  month = feb,
  journal = {The American Journal of Medicine},
  volume = {119},
  number = {2},
  pages = {166.e7-166.e16},
  issn = {00029343},
  doi = {10.1016/j.amjmed.2005.10.036},
  urldate = {2026-04-17},
  langid = {english}
}

@article{kimberlinValidityReliabilityMeasurement2008,
  title = {Validity and Reliability of Measurement Instruments Used in Research},
  author = {Kimberlin, Carole L. and Winterstein, Almut G.},
  year = 2008,
  month = dec,
  journal = {American Journal of Health-System Pharmacy},
  volume = {65},
  number = {23},
  pages = {2276--2284},
  issn = {1079-2082, 1535-2900},
  doi = {10.2146/ajhp070364},
  urldate = {2026-04-17},
  langid = {english}
}

@misc{salaudeenMeasurementMeaningValidityCentered2025,
  title = {Measurement to {{Meaning}}: {{A Validity-Centered Framework}} for {{AI Evaluation}}},
  shorttitle = {Measurement to {{Meaning}}},
  author = {Salaudeen, Olawale and Reuel, Anka and Ahmed, Ahmed and Bedi, Suhana and Robertson, Zachary and Sundar, Sudharsan and Domingue, Ben and Wang, Angelina and Koyejo, Sanmi},
  year = 2025,
  publisher = {arXiv},
  doi = {10.48550/arXiv.2505.10573},
  urldate = {2026-04-19},
  copyright = {Creative Commons Attribution Non Commercial No Derivatives 4.0 International}
}

@misc{wallachPositionEvaluatingGenerative2025,
  title = {Position: {{Evaluating Generative AI Systems Is}} a {{Social Science Measurement Challenge}}},
  shorttitle = {Position},
  author = {Wallach, Hanna and Desai, Meera and Cooper, A. Feder and Wang, Angelina and Atalla, Chad and Barocas, Solon and Blodgett, Su Lin and Chouldechova, Alexandra and Corvi, Emily and Dow, P. Alex and {Garcia-Gathright}, Jean and Olteanu, Alexandra and Pangakis, Nicholas and Reed, Stefanie and Sheng, Emily and Vann, Dan and Vaughan, Jennifer Wortman and Vogel, Matthew and Washington, Hannah and Jacobs, Abigail Z.},
  year = 2025,
  month = feb,
  journal = {arXiv.org},
  urldate = {2025-06-15},
  howpublished = {https://arxiv.org/abs/2502.00561v2},
  langid = {english}
}

@article{wangEvaluatingGeneralPurposeAI2026,
  title = {Evaluating {{General-Purpose AI}} with {{Psychometrics}}},
  author = {Wang, Xiting and Jiang, Liming and {Hern{\'a}ndez-Orallo}, Jos{\'e} and Stillwell, David and Chen, Shiqiang and Sun, Luning and Luo, Fang and Xie, Xing},
  year = 2026,
  month = apr,
  journal = {Communications of the ACM},
  pages = {3769688},
  issn = {0001-0782, 1557-7317},
  doi = {10.1145/3769688},
  urldate = {2026-04-19},
  langid = {english}
}

@INPROCEEDINGS{Beel2019e,   
author = {Beel, Joeran and Brunel, Victor},   
title = {Data Pruning in Recommender Systems Research: Best-Practice or Malpractice?},   
booktitle = {Proceedings of the 13th ACM Conference on Recommender Systems (RecSys)},   
year = {2019}, 
}

@article{harperMovieLensDatasetsHistory2015a,
  title = {The {{MovieLens Datasets}}: {{History}} and {{Context}}},
  shorttitle = {The {{MovieLens Datasets}}},
  author = {Harper, F. Maxwell and Konstan, Joseph A.},
  year = 2015,
  month = dec,
  journal = {ACM Trans. Interact. Intell. Syst.},
  volume = {5},
  number = {4},
  pages = {19:1--19:19},
  issn = {2160-6455},
  doi = {10.1145/2827872},
  urldate = {2026-04-22}
}

@book{voorhees2005trec,
  title={TREC: Experiment and evaluation in information retrieval},
  author={Voorhees, Ellen M and Harman, Donna K and others},
  volume={63},
  year={2005},
  publisher={MIT press Cambridge}
}

@article{luEffectPoolingEvaluation2016,
  title = {The Effect of Pooling and Evaluation Depth on {{IR}} Metrics},
  author = {Lu, Xiaolu and Moffat, Alistair and Culpepper, J. Shane},
  year = 2016,
  month = aug,
  journal = {Information Retrieval Journal},
  volume = {19},
  number = {4},
  pages = {416--445},
  issn = {1386-4564, 1573-7659},
  doi = {10.1007/s10791-016-9282-6},
  urldate = {2026-04-22},
  langid = {english}
}
